\documentstyle[prd,aps,preprint,tighten,epsfig]{revtex}

\begin{document}

\draft

\title{No Hope to Kinematically Detect the Effective Masses \\
of Muon and Tau Neutrinos}
\author{{\bf Zhi-zhong Xing}}
\address{Institute of High Energy Physics, Chinese Academy of Sciences, \\
P.O. Box 918 (4), Beijing 100039, China \\
({\it Electronic address: xingzz@mail.ihep.ac.cn}) } 
\maketitle

\begin{abstract}
We show that the recent WMAP data can impose a generous upper bound
on the effective masses of electron, muon and tau neutrinos defined 
in the kinematic measurements:
$\langle m\rangle^2_e + \langle m\rangle^2_\mu +
\langle m\rangle^2_\tau = m^2_1 + m^2_2 + m^2_3 < 0.5 ~ {\rm eV}^2$, 
or $\langle m\rangle_\alpha < 0.71$ eV (for $\alpha = e, \mu, \tau$).
When current neutrino oscillation data are taken into account,
we obtain $\langle m\rangle_e <0.24$ eV and 
$\langle m\rangle_\mu \approx \langle m\rangle_\tau < 0.24$ eV.
Thus there is no hope to kinematically detect
$\langle m\rangle_\mu$ and $\langle m\rangle_\tau$ in any realistic
experiments.
\end{abstract}


Thanks to the recent Super-Kamiokande \cite{SK}, K2K \cite{K2K}, 
SNO \cite{SNO} and KamLAND \cite{KM} experiments, we are
now convinced that the deficit of atmospheric $\nu_\mu$ neutrinos
and the deficit of solar $\nu_e$ neutrinos are both due to 
neutrino oscillations, a quantum phenomenon which can naturally
happen if neutrinos are massive and lepton flavors are mixed.
Furthermore, current experimental data indicate that solar and
atmospheric neutrino oscillations are dominated respectively by
$\nu_e\rightarrow \nu_\mu$ and $\nu_\mu\rightarrow \nu_\tau$ 
transitions. The neutrino mass-squared differences associated 
with solar and atmospheric neutrino oscillations are thus
defined as
\begin{eqnarray}
\Delta m^2_{\rm sun} & \equiv & \left | m^2_2 ~ - ~ m^2_1 \right | \; ,
\nonumber \\
\Delta m^2_{\rm atm} & \equiv & \left | m^2_3 ~ - ~ m^2_2 \right | \; ,
\end{eqnarray}
where $m_i$ (for $i=1,2,3$) denote the physical masses of three
neutrinos. Although a strong hierarchy between
$\Delta m^2_{\rm sun}$ and $\Delta m^2_{\rm atm}$ has been 
observed, the absolute values of $m_1$, $m_2$ and $m_3$ remain 
unknown. The kinematic limits on the effective masses of electron, 
muon and tau neutrinos can be obtained from the tritium $\beta$-decay 
$^3_1{\rm H} \rightarrow$$^3_2{\rm He} + e^- + \overline{\nu}_e$,
the $\pi^+ \rightarrow \mu^+ + \nu_\mu$ decay and the
$\tau \rightarrow 5\pi + \nu_\tau$ 
(or $\tau \rightarrow 3\pi + \nu_\tau$) decay, respectively. Today's
results are \cite{PDG02}
\begin{eqnarray}
\langle m\rangle_e & < & 3 ~ {\rm eV} \; ,
\nonumber \\
\langle m\rangle_\mu & < & 0.19 ~ {\rm MeV} \; ,
\nonumber \\
\langle m\rangle_\tau & < & 18.2 ~ {\rm MeV} \; .
\end{eqnarray}
One can see that the experimental sensitivity for  
$\langle m\rangle_\mu$ is more than four orders of magnitude
smaller than that for $\langle m\rangle_e$, and the experimental 
sensitivity for $\langle m\rangle_\tau$ is two orders of
magnitude lower than that for $\langle m\rangle_\mu$ \cite{Vogel}.
Is there any hope to detect $\langle m\rangle_\mu$ and 
$\langle m\rangle_\tau$ or to constrain them to a meaningful 
level of sensitivity? The answer to this question relies on how 
small $\langle m\rangle_\mu$ and $\langle m\rangle_\tau$ are. 

The main purpose of this short note is to calculate $\langle m\rangle_e$, 
$\langle m\rangle_\mu$ and $\langle m\rangle_\tau$ with the help
of the recent WMAP data \cite{WMAP} and neutrino oscillation data.
While $\langle m\rangle_e$ has been extensively studied in the 
literature \cite{Giunti}, a detailed analysis of $\langle m\rangle_\mu$ 
and $\langle m\rangle_\tau$ has not been done. It is therefore 
worthwhile to work out the upper bounds on $\langle m\rangle_\mu$ and 
$\langle m\rangle_\tau$ from currently available data, such that
one may definitely answer the afore-raised question. Independent of
the neutrino oscillation data, a generous upper limit 
$\langle m\rangle^2_e + \langle m\rangle^2_\mu +
\langle m\rangle^2_\tau = m^2_1 + m^2_2 + m^2_3 <
0.5 ~ {\rm eV}^2$ can be achieved from the
WMAP observation. Hence $\langle m\rangle_\alpha < 0.71$ eV holds for
$\alpha = e, \mu$ or $\tau$. Such an upper bound for 
$\langle m\rangle_\alpha$ will be reduced by a factor of three, if
current neutrino oscillation data are taken into account. In 
particular, $\langle m\rangle_\mu \approx \langle m\rangle_\tau$
is a good approximation and both of them are insensitive to the 
Dirac $CP$-violating phase of the lepton flavor mixing matrix. We
conclude that there is no hope to kinematically detect the effective
masses of muon and tau neutrinos.

\vspace{0.3cm}

Direct neutrino mass measurements are based on the analysis of
the kinematics of charged particles produced together with
neutrino flavor eigenstates $|\nu_\alpha\rangle$ 
(for $\alpha =e,\mu,\tau$), which are superpositions of neutrino
mass eigenstates $|\nu_i\rangle$ (for $i=1,2,3$):
\begin{equation}
\left ( \matrix{
\nu_e \cr
\nu_\mu \cr
\nu_\tau \cr} \right ) =
\left ( \matrix{
V_{e1}		& V_{e2}	& V_{e3} \cr
V_{\mu 1}	& V_{\mu 2}	& V_{\mu 3} \cr
V_{\tau 1}	& V_{\tau 2} 	& V_{\tau 3} \cr} \right )
\left ( \matrix{
\nu_1 \cr
\nu_2 \cr
\nu_3 \cr} \right ) \; .
\end{equation}
The unitary matrix $V$ is just the lepton flavor mixing matrix.
The effective masses of electron, muon and tau neutrinos in the
kinematic measurements can then be defined \cite{Vogel}:
\begin{equation}
\langle m\rangle^2_\alpha \equiv  
|V_{\alpha 1}|^2 m^2_1 + |V_{\alpha 2}|^2 m^2_2 +
|V_{\alpha 3}|^2 m^2_3 \; ,
\end{equation}
for $\alpha =e,\mu$ and $\tau$. The unitarity of $V$ leads 
straightforwardly to a simple sum rule between  
$\langle m\rangle^2_\alpha$ and $m^2_i$:
\begin{equation}
\langle m\rangle^2_e + \langle m\rangle^2_\mu +
\langle m\rangle^2_\tau \; = \; m^2_1 + m^2_2 + m^2_3 \; .
\end{equation}
Note that this sum rule allows us to derive an upper bound on
$\langle m\rangle^2_\alpha$ independent of any neutrino oscillation
data. This point can clearly be seen from the inequality
\footnote{It is worthwhile to mention that the sum
$\overline{m}^2 \equiv m^2_1 + m^2_2 + m^2_3$ is a crucial quantity 
in the thermal leptogenesis mechanism \cite{FY}, because it 
controls an important class of washout processes \cite{Buch}.} 
\begin{equation}
m^2_1 ~ + ~ m^2_2 ~ + ~ m^2_3 \; < 
\left (m_1 + m_2 + m_3 \right )^2  ,
\end{equation}
in which the sum of three neutrino masses has well be constrained 
by the recent WMAP data \cite{WMAP}: 
$m_1 + m_2 + m_3 < 0.71$ eV at the $95\%$ confidence level. 
Therefore,
\begin{equation}
\langle m\rangle^2_e ~ + ~ \langle m\rangle^2_\mu ~ + ~
\langle m\rangle^2_\tau \;\; < \;\; 0.50 ~ {\rm eV}^2 \;\; . 
\end{equation}
This generous upper bound implies that 
$\langle m\rangle^2_\alpha < 0.50 ~ {\rm eV}^2$ or
$\langle m\rangle_\alpha < 0.71 ~ {\rm eV}$ holds for
$\alpha = e,\mu$ and $\tau$. Two comments are then in order.
\begin{enumerate}
\item    The cosmological upper bound of $\langle m\rangle_\mu$ 
is more than five orders of magnitude smaller than its kinematic
upper bound given in Eq. (2). In comparison, the upper limit of
$\langle m\rangle_\tau$ set by the WMAP data is more than seven
orders of magnitude smaller than its kinematic upper limit.
It seems hopeless to improve the sensitivity of the kinematic 
measurements to the level of 0.71 eV.
\item    The cosmological upper bound of $\langle m\rangle_e$ 
is about four times smaller than its kinematic upper bound given
in Eq. (2). The former may be accessible in the future KATRIN 
experiment \cite{KATRIN}, whose sensitivity is expected to be
about 0.35 eV.
\end{enumerate}
If current data on neutrino oscillations are taken into account,
however, more stringent upper limits can be obtained for the
effective neutrino masses $\langle m\rangle_e$, 
$\langle m\rangle_\mu$ and $\langle m\rangle_\tau$.

\vspace{0.3cm}

To see how $\langle m\rangle_\alpha$ may be related to the parameters 
of neutrino oscillations, we make use of Eq. (1) to express $m_1$ and 
$m_2$ in terms of $m_3$, $\Delta m^2_{\rm sun}$ and 
$\Delta m^2_{\rm atm}$. The results are 
\begin{eqnarray}
m_1 & = & \sqrt{m^2_3 + p \Delta m^2_{\rm atm} + 
q \Delta m^2_{\rm sun}} \;\; ,
\nonumber \\
m_2 & = & \sqrt{m^2_3 + p \Delta m^2_{\rm atm}} \;\; ,
\end{eqnarray}
where $p=\pm 1$ and $q=\pm 1$ stand for four possible patterns of the
neutrino mass spectrum. For example,
$p = q = +1$ corresponds to $m_1 > m_2 > m_3$, and so on.
The present solar neutrino oscillation data favor $q = -1$ or
$m_1 < m_2$ \cite{Fit}, but the sign of $p$ has not been fixed.
Substituting Eq. (8) into Eq. (4), we obtain
\begin{equation}
\langle m\rangle_\alpha \; =\; \sqrt{m^2_3 +
p \left (1 - |V_{\alpha 3}|^2 \right ) \Delta m^2_{\rm atm} +
q |V_{\alpha 1}|^2 \Delta m^2_{\rm sun}} \; \; .
\end{equation}
Taking account of $|V_{e3}|^2 \ll 1$ \cite{CHOOZ} and 
$\Delta m^2_{\rm sun} \ll \Delta m^2_{\rm atm}$, we arrive at the
following approximation for $\langle m\rangle_e$:
\begin{equation}
\langle m\rangle_e \; \approx \; \sqrt{m^2_3 + p
\Delta m^2_{\rm atm}} \;\; .
\end{equation}
Taking account of the observed (nearly) maximal mixing factor of
atmospheric neutrino oscillations \cite{SK}, which is equivalent to 
$|V_{\mu 3}| \approx |V_{\tau 3}| \approx 1/\sqrt{2}$ for
$|V_{e3}| \ll 1$, we obtain
\begin{eqnarray}
\langle m\rangle_\mu & \approx & \sqrt{m^2_3 + \frac{p}{2}
\Delta m^2_{\rm atm}} \;\; ,
\nonumber \\
\langle m\rangle_\tau & \approx & \sqrt{m^2_3 + \frac{p}{2}
\Delta m^2_{\rm atm}} \;\; .
\end{eqnarray} 
One can see that $\langle m\rangle_\mu \approx \langle m\rangle_\tau$
is a natural consequence of current neutrino oscillation data.
In addition, Eqs. (10) and (11) tell us that $\langle m\rangle_e$ is 
slightly larger than $\langle m\rangle_\mu$ and $\langle m\rangle_\tau$ 
for $p=+1$ or $m_2 > m_3$; and it is slightly smaller than 
$\langle m\rangle_\mu$ and $\langle m\rangle_\tau$ for $p=-1$ or
$m_2 < m_3$.

In general, one may analyze the correlation between 
$\langle m\rangle_\alpha$ and $m_3$ by use of Eq. (9), only if 
$|V_{\alpha 1}|$ and $|V_{\alpha 3}|$ are already measured.
Since the mixing angles of solar, atmospheric
and CHOOZ (or Palo Verde) reactor \cite{CHOOZ} neutrino oscillations
read as
\begin{eqnarray}
\sin^2 2\theta_{\rm sun} & = & 4 |V_{e1}|^2 |V_{e2}|^2 \; ,
\nonumber \\
\sin^2 2\theta_{\rm atm} & = & 4 |V_{\mu 3}|^2
\left ( 1 - |V_{\mu 3}|^2 \right ) \; ,
\nonumber \\
\sin^2 2\theta_{\rm chz} & = & 4 |V_{e3}|^2
\left ( 1 - |V_{e3}|^2 \right ) \; ,
\end{eqnarray}
we reversely obtain \cite{Xing02}
\begin{eqnarray}
|V_{e1}| & = & \frac{1}{\sqrt 2} \sqrt{ \cos^2\theta_{\rm chz} +
\sqrt{\cos^4\theta_{\rm chz} - \sin^2 2\theta_{\rm sun}}} \;\; ,
\nonumber \\
|V_{e2}| & = & \frac{1}{\sqrt 2} \sqrt{ \cos^2\theta_{\rm chz} -
\sqrt{\cos^4\theta_{\rm chz} - \sin^2 2\theta_{\rm sun}}} \;\; ,
\nonumber \\
|V_{e3}| & = & \sin\theta_{\rm chz} \; ,
\nonumber \\
|V_{\mu 3}| & = & \sin\theta_{\rm atm} \; ,
\nonumber \\
|V_{\tau 3}| & = & \sqrt{\cos^2\theta_{\rm chz} -
\sin^2\theta_{\rm atm}} \;\; .
\end{eqnarray}
The other four matrix elements of $V$ (i.e., $|V_{\mu 1}|$, $|V_{\mu 2}|$, 
$|V_{\tau 1}|$ and $|V_{\tau 2}|$) are unable to be determined from the
afore-mentioned neutrino oscillation experiments. They can be derived
from Eq. (13), however, if the Dirac $CP$-violating phase in the standard
parametrization of $V$ \cite{FX01} is taken into account. 
The explicit expressions of $|V_{\mu 1}|$, $|V_{\mu 2}|$,
$|V_{\tau 1}|$ and $|V_{\tau 2}|$ are given in Appendix A.
We see that $\langle m\rangle_\mu$ and $\langle m\rangle_\tau$ 
depend on the Dirac phase $\delta$ in the chosen 
parametrization of $V$. However, the sensitivity of 
$\langle m\rangle_\mu$ or $\langle m\rangle_\tau$ to $\delta$
is negligibly weak. The reason is simply that the contribution of
$\cos\delta$ to $|V_{\mu 1}|$ (or $|V_{\tau 1}|$) is suppressed
by $|V_{e3}|$ or $\sin\theta_{\rm chz}$, and the contribution of 
$|V_{\mu 1}|$ (or $|V_{\tau 1}|$) to $\langle m\rangle_\mu$ 
(or $\langle m\rangle_\tau$) is further suppressed by 
$\Delta m^2_{\rm sun}$. It is therefore hopeless to probe the
$CP$-violating phase $\delta$, even if $\langle m\rangle_\mu$ and 
$\langle m\rangle_\tau$ could be measured.

To numerically determine the upper bound of $\langle m\rangle_\alpha$, 
we need first of all work out the upper limit of $m_3$ set by
the WMAP and neutrino oscillation data. In FIG. 1, we show the 
dependence of $m_1 + m_2 + m_3$ on $m_3$, where the best-fit
values $\Delta m^2_{\rm sun} = 7.3 \times 10^{-5} ~ {\rm eV}^2$
and $\Delta m^2_{\rm atm} = 2.5 \times 10^{-3} ~ {\rm eV}^2$ 
\cite{Fit} have typically been input. Note that only the 
$m_1 < m_2$ case, which is supported by current solar neutrino
oscillation data, is taken into account. For the $m_2 < m_3$
case, $m_3$ has an lower bound 
$m_3 \geq \sqrt{\Delta m^2_{\rm atm} + \Delta m^2_{\rm sun}} 
\approx 0.051$ eV; but for the $m_2 > m_3$ case, even $m_3 =0$ is 
allowed (inverted hierarchy). We see that these two cases become 
indistinguishable for $m_3 \geq 0.2$ eV, implying the near 
degeneracy of three neutrino masses. Once the WMAP limit
$m_1 + m_2 + m_3 < 0.71$ eV is included, we immediately get 
$m_3 < 0.24$ eV. Then we have $m_i < 0.24$ eV for 
$i=1,2$ or $3$. This result is apparently 
consistent with those obtained in Ref. \cite{MAP}. 

Next we evaluate $\langle m\rangle_\alpha$ by using the best-fit
values $\theta_{\rm sun} \approx 33^\circ$ and
$\theta_{\rm atm} \approx 45^\circ$ \cite{Fit} in addition to taking 
$\theta_{\rm chz} \approx 5^\circ$ as a typical input, which is
compatible with $\sin^2 2\theta_{\rm chz} <0.1$ extracted from 
the CHOOZ reactor neutrino experiment \cite{CHOOZ}
\footnote{Note that $\theta_{\rm chz} \sim 5^\circ$ is also favored 
in a number of phenomenological models of lepton mass matrices. 
See Ref. \cite{Review} for a review with extensive references.}.
Our numerical results for $\langle m\rangle_e$,
$\langle m\rangle_\mu$ and $\langle m\rangle_\tau$ changing
with $m_3$ are shown in FIG. 2. Once again, it is impossible
to distinguish between the $m_2 < m_3$ case and the 
$m_2 > m_3$ case for $m_3 \geq 0.2$ eV. We find that the
dependence of $\langle m\rangle_\alpha$ on $m_3$ is very
similar to the dependence of $m_1 + m_2 + m_3$ on $m_3$.
As expected in Eqs. (10) and (11), 
$\langle m\rangle_e > \langle m\rangle_\mu \approx 
\langle m\rangle_\tau$ holds for $m_3 < m_2$ (curves a and
b in FIG. 2); and 
$\langle m\rangle_e < \langle m\rangle_\mu \approx 
\langle m\rangle_\tau$ holds for $m_3 > m_2$ (curves c and d
in FIG. 2). 
In view of the upper limit $m_3 \leq 0.24$ eV obtained
above, we arrive at $\langle m\rangle_\alpha \leq 0.24$ eV for
$\alpha = e, \mu$ or $\tau$. This upper bound is suppressed by
a factor of three, compared to the upper bound obtained from
Eq. (7) which is independent of the neutrino oscillation data.

The result 
$\langle m\rangle_\mu \approx \langle m\rangle_\tau < 0.24$ eV
implies that there is no hope to kinematically detect the 
effective masses of muon and tau neutrinos. As the WMAP upper
bound is in general valid for a sum of the masses of all 
possible relativistic neutrinos (no matter whether they are 
active or sterile), it seems unlikely to loosen the upper 
limit of $\langle m\rangle_\alpha$ obtained above in the
assumption of only active neutrinos. Therefore, the kinematic
measurements of $\langle m\rangle_\mu$ and $\langle m\rangle_\tau$
have little chance to reveal the existence of any exotic 
neutral particles with masses much larger than the light 
neutrino masses. 

\vspace{0.3cm}

The author is grateful to W.L. Guo for some discussions. This work 
was supported in part by National Natural Science Foundation of China.

\newpage

\appendix
\section{}

This Appendix is devoted to the calculation of $|V_{\mu 1}|$, 
$|V_{\mu 2}|$, $|V_{\tau 1}|$ and $|V_{\tau 2}|$, which are entirely 
unrestricted by current neutrino oscillation data.
Taking account of the Dirac $CP$-violating phase in the standard 
parametrization of $V$ \cite{FX01}, we find
\begin{eqnarray}
|V_{\mu 1}| & = & \frac{\sqrt{|V_{e2}|^2 |V_{\tau 3}|^2
+ |V_{e1}|^2 |V_{e3}|^2 |V_{\mu 3}|^2 +
2 |V_{e1}| |V_{e2}| |V_{e3}| |V_{\mu 3}| |V_{\tau 3}| \cos\delta}}
{1 - |V_{e3}|^2} \;\; ,
\nonumber \\ \nonumber \\
|V_{\mu 2}| & = & \frac{\sqrt{|V_{e1}|^2 |V_{\tau 3}|^2
+ |V_{e2}|^2 |V_{e3}|^2 |V_{\mu 3}|^2 -
2 |V_{e1}| |V_{e2}| |V_{e3}| |V_{\mu 3}| |V_{\tau 3}| \cos\delta}}
{1 - |V_{e3}|^2} \;\; ,
\nonumber \\ \nonumber \\
|V_{\tau 1}| & = & \frac{\sqrt{|V_{e2}|^2 |V_{\mu 3}|^2
+ |V_{e1}|^2 |V_{e3}|^2 |V_{\tau 3}|^2 -
2 |V_{e1}| |V_{e2}| |V_{e3}| |V_{\mu 3}| |V_{\tau 3}| \cos\delta}}
{1 - |V_{e3}|^2} \;\; ,
\nonumber \\ \nonumber \\
|V_{\tau 2}| & = & \frac{\sqrt{|V_{e1}|^2 |V_{\mu 3}|^2
+ |V_{e2}|^2 |V_{e3}|^2 |V_{\tau 3}|^2 +
2 |V_{e1}| |V_{e2}| |V_{e3}| |V_{\mu 3}| |V_{\tau 3}| \cos\delta}}
{1 - |V_{e3}|^2} \;\; .
\end{eqnarray}
The explicit expressions of $|V_{\mu 1}|$, $|V_{\mu 2}|$,
$|V_{\tau 1}|$ and $|V_{\tau 2}|$ in terms of $\theta_{\rm sun}$,
$\theta_{\rm atm}$, $\theta_{\rm chz}$ and $\delta$ can then
be obtained from Eq. (13) and Eq. (A1):
\begin{eqnarray}
|V_{\mu 1}| & = & \left [ \frac{\cos^2\theta_{\rm atm}}{2} ~ - ~
\frac{\cos^2\theta_{\rm chz} - \sin^2\theta_{\rm atm} 
\left (1 + \sin^2\theta_{\rm chz} \right )}{2\cos^4\theta_{\rm chz}}
\sqrt{\cos^4\theta_{\rm chz} - \sin^2 2\theta_{\rm sun}}
\right .
\nonumber \\
& & \left . + ~ \frac{\sin 2\theta_{\rm sun} \sin\theta_{\rm atm} 
\sin\theta_{\rm chz} \cos\delta}{\cos^4\theta_{\rm chz}}
 \sqrt{\cos^2\theta_{\rm chz} - \sin^2\theta_{\rm atm}} 
\right ]^{1/2} \; ,
\end{eqnarray}
\begin{eqnarray}
|V_{\mu 2}| & = & \left [ \frac{\cos^2\theta_{\rm atm}}{2} ~ + ~
\frac{\cos^2\theta_{\rm chz} - \sin^2\theta_{\rm atm} 
\left (1 + \sin^2\theta_{\rm chz} \right )}{2\cos^4\theta_{\rm chz}}
\sqrt{\cos^4\theta_{\rm chz} - \sin^2 2\theta_{\rm sun}}
\right .
\nonumber \\
& & \left . - ~ \frac{\sin 2\theta_{\rm sun} \sin\theta_{\rm atm} 
\sin\theta_{\rm chz} \cos\delta}{\cos^4\theta_{\rm chz}}
 \sqrt{\cos^2\theta_{\rm chz} - \sin^2\theta_{\rm atm}} 
\right ]^{1/2} \; ,
\end{eqnarray}
\begin{eqnarray}
|V_{\tau 1}| & = & \left [ \frac{\sin^2\theta_{\rm atm} +
\sin^2\theta_{\rm chz}}{2} ~ - ~
\frac{4 \sin^2\theta_{\rm atm} \left (1 + \sin^2\theta_{\rm chz} \right )
- \sin^2 2\theta_{\rm chz}}{8\cos^4\theta_{\rm chz}}
\sqrt{\cos^4\theta_{\rm chz} - \sin^2 2\theta_{\rm sun}}
\right .
\nonumber \\
& & \left . - ~ \frac{\sin 2\theta_{\rm sun} \sin\theta_{\rm atm} 
\sin\theta_{\rm chz} \cos\delta}{\cos^4\theta_{\rm chz}}
 \sqrt{\cos^2\theta_{\rm chz} - \sin^2\theta_{\rm atm}} 
\right ]^{1/2} \; ,
\end{eqnarray}
\begin{eqnarray}
|V_{\tau 2}| & = & \left [ \frac{\sin^2\theta_{\rm atm} +
\sin^2\theta_{\rm chz}}{2} ~ + ~
\frac{4 \sin^2\theta_{\rm atm} \left (1 + \sin^2\theta_{\rm chz} \right )
- \sin^2 2\theta_{\rm chz}}{8\cos^4\theta_{\rm chz}}
\sqrt{\cos^4\theta_{\rm chz} - \sin^2 2\theta_{\rm sun}}
\right .
\nonumber \\
& & \left . + ~ \frac{\sin 2\theta_{\rm sun} \sin\theta_{\rm atm} 
\sin\theta_{\rm chz} \cos\delta}{\cos^4\theta_{\rm chz}}
 \sqrt{\cos^2\theta_{\rm chz} - \sin^2\theta_{\rm atm}} 
\right ]^{1/2} \; .
\end{eqnarray}
These results are useful for a numerical analysis of $V$ by use of
current experimental data on neutrino oscillations, if $\delta$ is
allowed to vary from 0 to $\pi$.

\newpage

\begin{figure}
\vspace{-0.2cm}
\epsfig{file=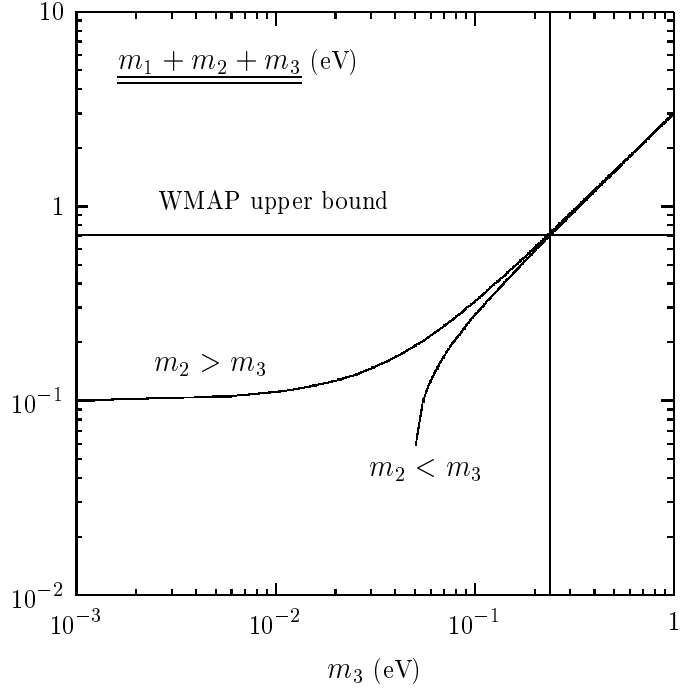,bbllx=0.5cm,bblly=8cm,bburx=17cm,bbury=30cm,%
width=15.5cm,height=22cm,angle=0,clip=}
\vspace{-9.3cm}
\caption{Illustrative dependence of $m_1 + m_2 + m_3$ on $m_3$.
The WMAP result sets an upper limit on $m_3$; i.e.,
$m_3 < 0.24$ eV for both $m_3 > m_2$ and $m_3 < m_2$ cases.}
\end{figure}

\begin{figure}
\vspace{-0.2cm}
\epsfig{file=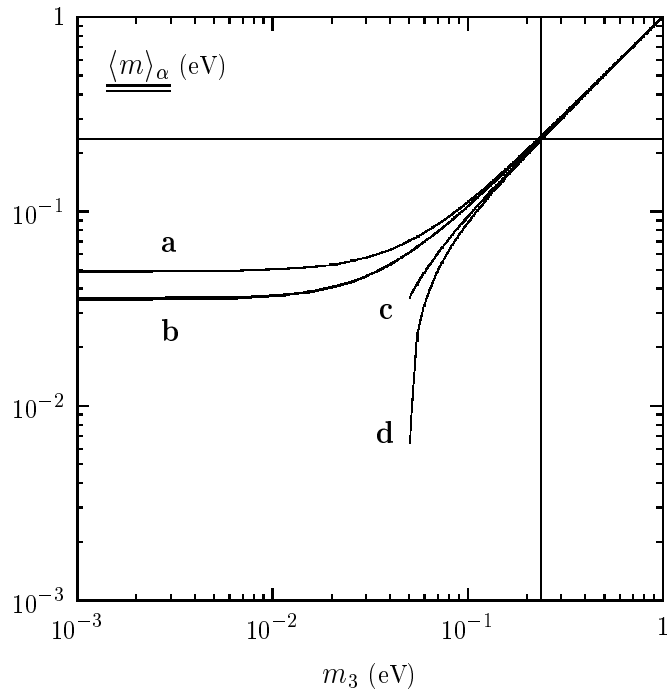,bbllx=0.5cm,bblly=8cm,bburx=17cm,bbury=30cm,%
width=15.5cm,height=22cm,angle=0,clip=}
\vspace{-9.3cm}
\caption{Illustrative dependence of $\langle m\rangle_\alpha$ 
(for $\alpha = e, \mu, \tau$)
on $m_3$. Curve {\bf a}: $\langle m\rangle_e$ with $m_3 < m_2$;
Curve {\bf b}: $\langle m\rangle_\mu \approx \langle m\rangle_\tau$ 
with $m_3 < m_2$; 
Curve {\bf c}: $\langle m\rangle_\mu \approx \langle m\rangle_\tau$ 
with $m_3 > m_2$; and
Curve {\bf d}: $\langle m\rangle_e$ with $m_3 > m_2$. The WMAP
result leads to $\langle m\rangle_\alpha <0.24$ eV.}
\end{figure}
\end{document}